\documentclass[twocolumn]{aastex6}

\usepackage{amsmath}
\usepackage{amssymb}
\usepackage{amsthm}
\usepackage{array}
\usepackage{float}
\usepackage{graphicx}
\usepackage{subfigure}
\usepackage{color}
\usepackage{hyperref}

\newcounter{ichi}
\newcounter{ni}
\newcounter{san}
\newcounter{yon}
\def\be{\begin{equation}}
\def\ee{\end{equation}}
\def\ba{\begin{eqnarray}}
\def\ea{\end{eqnarray}}

\makeatletter
\newcommand\mytuple[1]{%
  \@tempcnta=0
  \bigl\langle
  \@for\@ii:=#1\do{%
    \@insertbreakingcomma
    \textit{\@ii}
  }%
  \bigr\rangle
}
\def\@insertbreakingcomma{%
  \ifnum \@tempcnta = 0 \else\,,\ \linebreak[1] \fi
  \advance\@tempcnta\@ne
}
\makeatother

\preprint{}
\shorttitle{Jetted Electromagnetic Counterparts of Supermassive Black Hole Coalescences}

\begin{document}
\title{Post-Merger Jets from Supermassive Black Hole Coalescences as Electromagnetic Counterparts of Gravitational Wave Emission}
\author{Chengchao Yuan\altaffilmark{1,2,3}}\email{cxy52@psu.edu}
\author{Kohta Murase\altaffilmark{1,2,3,4}}\email{murase@psu.edu}
\author{B. Theodore Zhang\altaffilmark{1,2,3}}
\author{Shigeo S. Kimura\altaffilmark{5}}
\author{Peter M\'esz\'aros\altaffilmark{1,2,3}}
\altaffiltext{1}{Department of Physics, The Pennsylvania State University, University Park, PA 16802, USA}
\altaffiltext{2}{Department of Astronomy \& Astrophysics, The Pennsylvania State University, University Park, PA 16802, USA}
\altaffiltext{3}{Center for Multimessenger Astrophysics, Institute for Gravitation and the Cosmos, The Pennsylvania State University, University Park, PA 16802, USA}
\altaffiltext{4}{Center for Gravitational Physics, Yukawa Institute for Theoretical Physics, Kyoto University, Kyoto, Kyoto 606-8502, Japan}
\altaffiltext{5}{Frontier Research Institute for Interdisciplinary Sciences; Astronomical Institute, Tohoku University, Sendai 980-8578, Japan}

\date{\today}
\begin{abstract}
As a powerful source of gravitational waves (GW), a supermassive black hole (SMBH) merger may be accompanied by a relativistic jet that leads to detectable electromagnetic (EM) emission. We model the propagation of post-merger jets inside a pre-merger wind bubble formed by disk-driven winds, and calculate multi-wavelength EM spectra from the forward shock region. 
We show that the non-thermal EM signals from SMBH mergers are detectable up to the detection horizon of future GW facilities such as the Laser Interferometer Space Antenna (LISA). 
Calculations based on our model predict slowly fading transients with time delays from days to months after the coalescence,  leading to implications for EM follow-up observations after the GW detection. 
\end{abstract}

\maketitle

\section{\label{sec:introduction}Introduction}
Supermassive black hole (SMBH) mergers are ubiquitous in the history of the Universe \citep{begelman1980massive,kormendy2013coevolution,2020MNRAS.498.5652K} and can produce powerful gravitational wave (GW) bursts when they coalesce \citep[e.g., ][]{thorne1976gravitational,2004ApJ...611..623S}, making them promising candidates for GW detectors such as Laser Interferometer Space Antenna \citep[LISA,][]{amaro2017laser,baker2019space} and pulsar timing arrays \citep[PTAs, e.g.,][]{2017NatAs...1..886M,taylor2019supermassive,2020ApJ...905L..34A} in single-source and/or stochastic GW background searches.
The accretion activity between the binary system and the surrounding disk can produce multi-wavelength electromagnetic (EM) emission \citep[e.g., ][]{milosavljevic2005afterglow,2012ApJ...749L..32M,farris2015binary,2017PhRvD..96l3003K,haiman2017electromagnetic,d2018electromagnetic}, and the time-variable EM signatures from the circumbinary disks could be detectable \citep[e.g.,][]{schnittman2008infrared,haiman2009identifying,tanaka2010time}.
The spinning SMBH expected to form after the SMBHs have coalesced may also lead to relativistic jets, in which particle acceleration will take place. 
The resulting non-thermal emission from the accelerated electrons may provide a promising post-merger EM counterpart of the GW emission, and will not only provide complementary information on SMBH mergers but also shed light on the physical processes in these systems \citep[e.g.,][]{2011CQGra..28i4021S,2018ASPC..517..781R,meszaros2019multi}. \cite{yuan2020high} recently suggested that the SMBH mergers can also be high-energy neutrino emitters, and demonstrated that they are also promising targets for high-energy multi-messenger astrophysics~\citep{Murase:2019tjj}.

We study the EM emission produced in relativistic jets launched after the coalescence of SMBHs. The physical picture is that the disk winds originating from the circumbinary disk and mini-disks around each SMBH form a pre-merger wind bubble, and jets powered by the Blandford-Znajek \citep[BZ,][]{blandford1977electromagnetic} mechanism are launched after the merger. The jets push ahead inside the pre-merger disk wind material, resulting in the formation of forward and reverse shocks. In the forward shock region, electrons are accelerated to high energies with a power-law distribution as observed in afterglows of gamma-ray bursts (GRBs) \citep[e.g.,][]{Meszaros:2006rc}. These particles then produce broadband non-thermal EM emission through synchrotron and synchrotron self-Compton (SSC) processes. 

This letter is organized as follows. In Sec.~\ref{sec:dynamics}, we introduce the physical conditions of the pre-merger wind bubble and model the propagation of jets. The radiation processes and the resulting photon spectra, light curves and detection horizons are presented in Sec.~\ref{sec:emissions}. In Sec. \ref{sec:discussion}, we discuss implications of our results. Throughout the letter, we use the conventional notation $Q_x=Q/10^x$ and physical quantities are written in the centimeter-gram-second units, unless otherwise specified.

\section{Jet dynamics}\label{sec:dynamics}
We discuss here the physical conditions in a pre-merger circumbinary environment and derive relevant quantities that describe the jet propagation. We consider on-axis observers, which is sufficient for the purpose of this work. The emission region is typically expected to be only mildly relativistic on time scales of interest (the corresponding observation time after the jet launch is $T\sim10^5-10^6$~s). 

\begin{figure*}\centering
\includegraphics[width=0.45\textwidth]{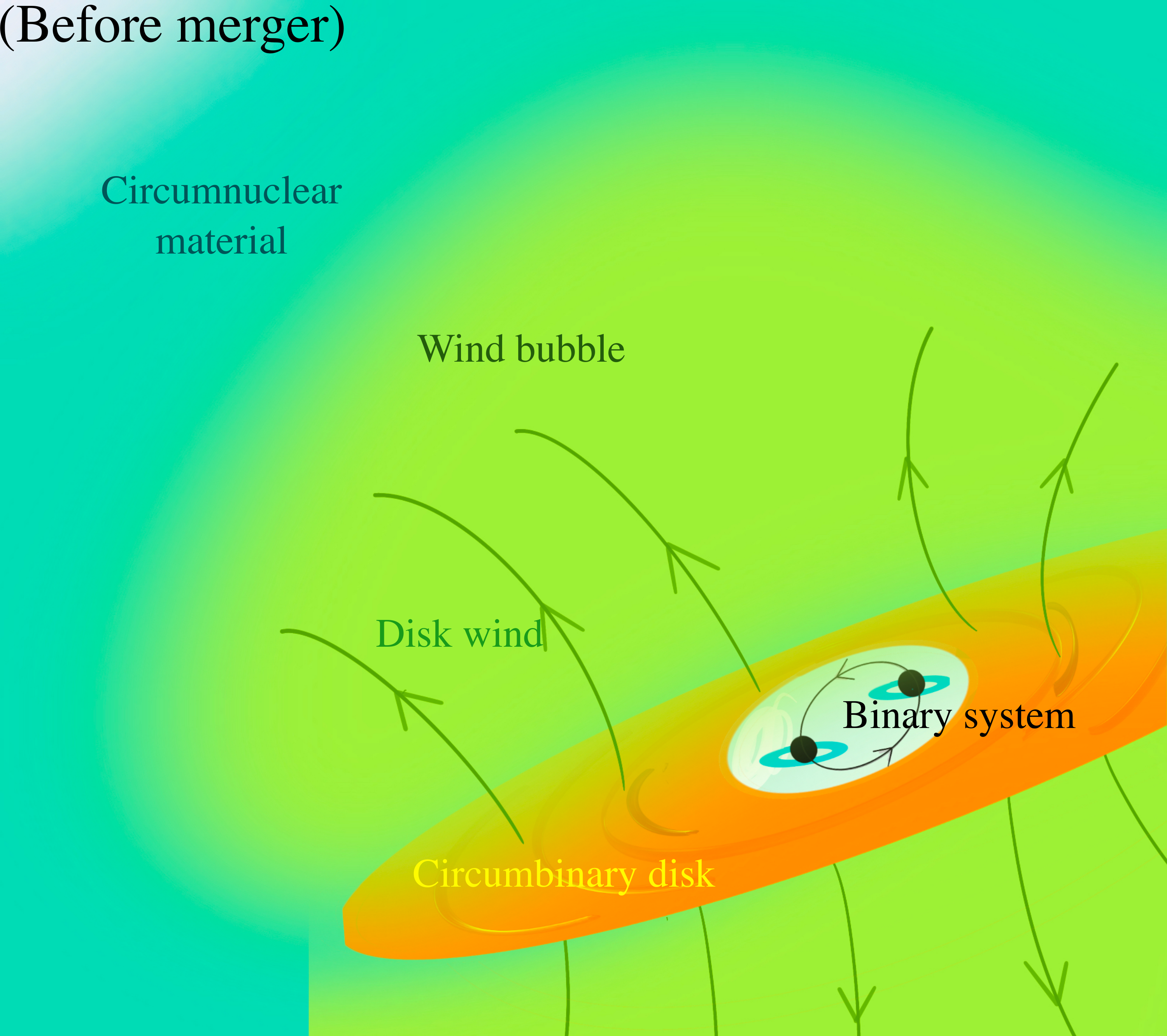}
\includegraphics[width=0.45\textwidth]{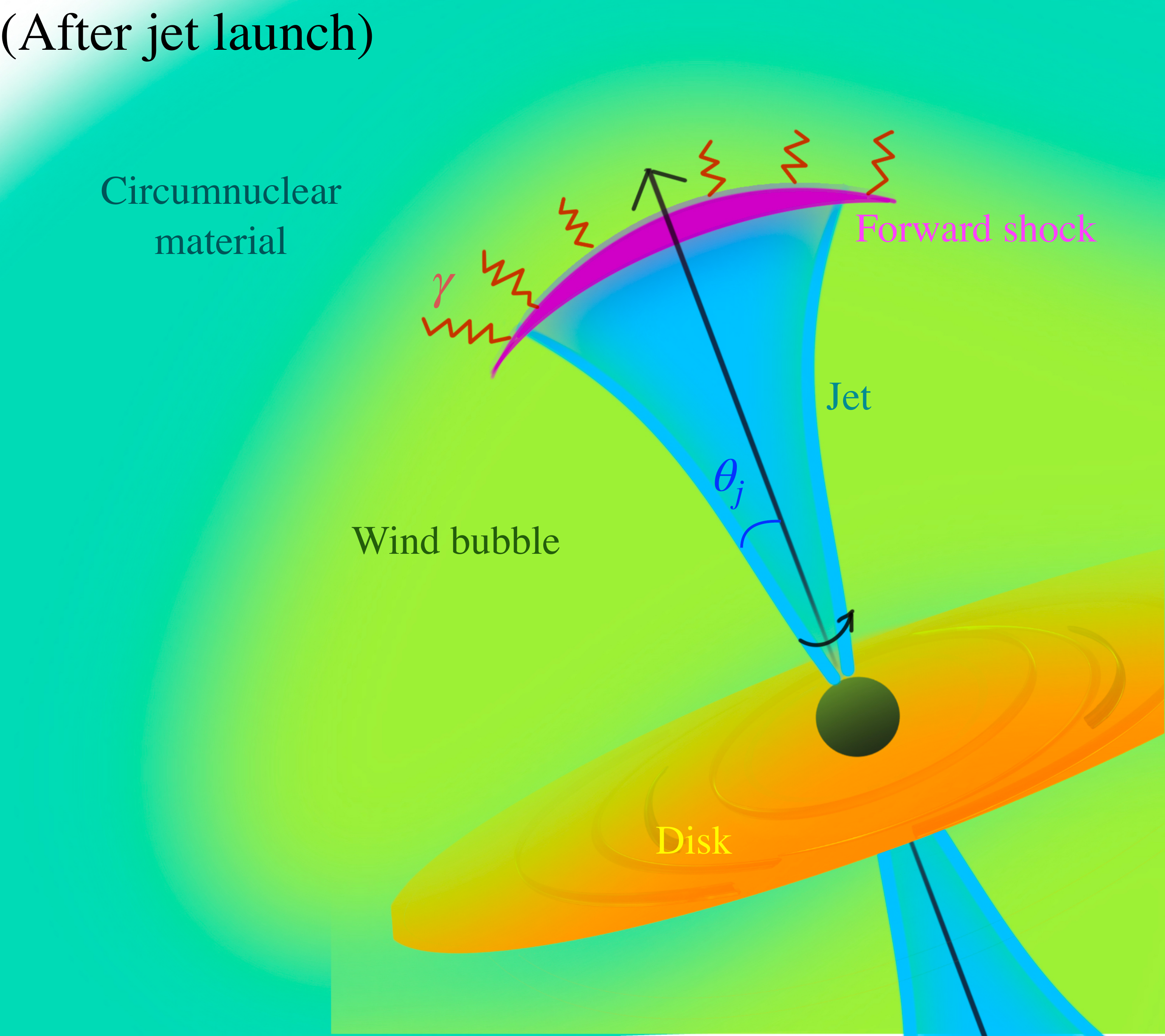}
\caption{Schematic description of our model. {\bf Left panel:} pre-merger disk winds launched from the circumbinary disk. The green arrows illustrate the disk-driven outflows that form a wind bubble. Mini-disks around each SMBH are also shown. {\bf Right panel:} post-merger jets launched by a merged SMBH. The forward shock region is shown as the purple area. The cocoon is not depicted. }
\label{fig:schematic}
\end{figure*}

{{Numerical simulations have demonstrated that binary SMBH mergers can produce jet-like emissions driven by the Poynting outflow \citep[e.g.,][]{2017PhRvD..96l3003K}.}} We assume that a jet is launched after the coalescence and subsequently propagates in the wind bubble formed by pre-merger disk winds. {Fig.~\ref{fig:schematic} schematically illustrates the configuration of the system. The disk wind expands in the gaseous environment of the host galaxy. We focus on emissions from the shock between the jet and the wind bubble.} Initially, the circumbinary disk can react promptly to the evolution of the binary system. The ratio between the disk radius $R_d$ and the semi-major axis of the binary system $a$ remains unchanged ($R_d/a\sim2$), until the inspiral time scale $t_{\rm GW}$ of the binary system \citep[e.g.,][]{shapiro2008black} equals the viscosity time scale $t_{\rm vis}$ \citep[e.g.,][]{pringle1981accretion}, which is known as the disk decoupling. 
After the disk becomes decoupled, the merger of SMBHs in binary system occurs within the time interval $t_m\sim 3\times10^{-2}~{\rm yr}~M_{\rm BH,6}\alpha_{-1}^{-8/5}h_{-1}^{-16/5}$, where $M_{\rm BH}=10^6~M_{\rm BH,6}M_{\odot}$ is the mass of the binary system, the dimensionless parameter $h$ is defined as $h=H/R_d$, $\alpha\sim0.1$ is the viscosity parameter, and $H$ is the disk scale height. The disk gas starts to fill the cavity between the disk and the SMBHs in the viscosity timescale, 
\begin{linenomath*}
\begin{equation}
t_{\rm vis}\sim0.1~{\rm yr}~M_{\rm BH,6}\alpha_{-1}^{-8/5}h_{-1}^{-16/5}
\label{eq:t_vis}
\end{equation}
\end{linenomath*}
after the coalescence \citep{farris2015binary}. This leads to a time delay ($t_{\rm delay}\sim t_{\rm vis}$) of days to months between the GW burst and the launch of post-merger jets, if $h\sim0.1-0.3$ is assumed. {However, for a thick and highly magnetized disk with $h\sim\alpha\sim1$, $t_{\rm delay}$ could be much shorter.}

On the other hand, within the duration of these two short-term processes, e.g., $t_{\rm m}$ and $t_{\rm vis}$, the disk wind radius may reach $v_{d}(t_{\rm vis}-t_m)\sim 10^{14}-10^{\rm 16}~\rm cm$ above the disk, where $v_d$ is the disk wind velocity that is of the order of the escape velocity, $v_{\rm esc}(R_{d,\rm dec})\approx\sqrt{2GM_{\rm BH}/R_{d,\rm dec}}$ for the circumbinary disk, and $R_{d,\rm dec}\approx1.2\times10^{13}~{\rm cm} ~M_{{\rm BH},6}\alpha_{-1}^{-2/5}h_{-1}^{-4/5}$ is the radius of the circumbinary disk at the decoupling. 
In reality, not only the circumbinary disk but also mini-disks around two SMBHs contribute, which would make the wind bubble more complicated. 
For simplicity, we assume the density profile of the winds at the decoupling to obtain the density distribution of the wind bubble at larger distances, 
\begin{linenomath*}
\begin{equation}
\varrho_w(r)=\frac{\eta_w(1+\chi)\dot M_{\rm BH}}{4\pi r^2v_{d}}\equiv Dr^{-2},
\label{eq:rho_w}
\end{equation}
\end{linenomath*}
where $\dot M_{\rm BH}$ is the mass accretion rate onto the binary system, $\chi\sim1$ is introduced to take into account the contribution of mini-disks, and $\eta_w$ represents the fraction of accreted mass converted to the disk wind. 
According to the simulations, for SANE (Standard And Normal Evolution) models, the parameter $\eta_w$ may vary from $10^{-4}$ to $10^{-1}$ \citep{jiang2019super,jiang2019global,ohsuga2009global} when the mass accretion rate changes from sub-Eddington to super-Eddington. 
In MAD (Magnetically Arrested Disk) models, $\eta_w$ can reach $10^{-2}$ to $10^{-1}$ \citep{akiyama2019first}. With $v_d\sim v_{\rm esc}(R_{d,\rm dec})$, 
we have $D\simeq5.9\times10^{11}~{\rm g~cm^{-1}}~\tilde\eta_{w,-1.5}(\dot{m}/0.5)M_{\rm BH,6}\beta_{d,-1}^{-1}$, where $\tilde{\eta}_w\equiv(1+\chi)\eta_w$, $\beta_{d,-1}\equiv v_d/(0.1c)$, the parameter $\dot m$ is defined as the ratio of $\dot M_{\rm BH}$, and the Eddington value $\dot M_{\rm Edd}\equiv 10L_{\rm Edd}/c^2$ (assuming a radiation efficiency of 0.1). 

After the coalescence, a powerful jet driven by the spin energy of the newly formed SMBH can appear, subsequently propagating in the pre-merger wind bubble. Considering a sub-Eddington accretion rate with the MAD configuration, we estimate the jet kinetic luminosity to be 
\begin{linenomath*}
\begin{eqnarray}
L_{k,j}&=&\eta_j\dot M_{\rm BH}c^2\nonumber\\
&\simeq&6.3\times10^{44}{~\rm erg\ s^{-1}}\eta_j(\dot m/0.5)M_{\rm BH,6},
\label{eq:L_k}
\end{eqnarray}
\end{linenomath*}
where $\eta_j\sim0.3-1$ is the ratio of the accretion energy converted to the jet energy \citep{tchekhovskoy2011efficient}. 

Following the standard jet propagation theory ~\citep[e.g.,][]{bromberg2011propagation,mizuta2013opening}, we write down the dimensionless parameter that represents the ratio of the energy density of the jet and the rest-mass energy density of the surrounding medium 
\begin{linenomath*}
\begin{equation}
\tilde L\approx\frac{L_{k,\rm iso}}{{4\pi r^2\varrho_wc^3}}\simeq 63~\tilde{\eta}_{w,-1.5}^{-1}\eta_j\theta_{j,-0.5}^{-2}\beta_{d,-1},
\end{equation}
\end{linenomath*}
where $\theta_j$ is the jet opening angle, $L_{k,\rm iso}\approx2L_{k,j}/\theta_j^2$ is the isotropic-equivalent luminosity. 
Since the quantity $\tilde L$ lies in the regime $\theta_j^{-4/3}\ll\tilde L$, we expect that the jet is 
``uncollimated'' for our fiducial parameters\footnote{However, jet collimation, which was assumed in \citet{yuan2020high}, would be achievable for the super-Eddington accretion accompanied by disk winds with $\eta_w\sim0.1-0.3$.}.
This situation is similar to that in choked jet propagation in the circumstellar material~\citep{Senno:2015tsn,Nakar:2015tma}, and $\beta_h$ is evaluated from $\tilde L$. 
In the relativistic limit, the jet head Lorentz factor is $\Gamma_h\approx\tilde{L}^{1/4}/\sqrt{2}$~\citep{Senno:2015tsn}, and we have $\Gamma_h\sim2$ in our fiducial case with $\tilde\eta_w=10^{-1.5}$. 
Note that the jet head radius is $R_h=c\beta_h\hat T\approx c\hat T$, and $\hat T$ is introduced to represent time measured in the central engine frame, which can be converted to the observation time $T$ via  $T=(1+z)(1-\beta_h)\hat T$ (that is $T\approx (1+z)\hat T/[2\Gamma_h^2]$ in the relativistic limit) for on-axis observers. 

Furthermore, to ensure particle acceleration, we impose radiation constraints requiring that the shock is collisionless, without being mediated by radiation~\citep{Murase:2013ffa,Senno:2015tsn}. 
Here, ignoring effects of pair production, we use the conservative condition, $\tau_T\approx\varrho_w\sigma_TR_h/m_p<1$,  where $\sigma_T$ is the Thomson cross section. Numerically, this condition is satisfied at $\hat T\gtrsim 10 ~\rm s$, which is much shorter than the duration of EM emission.

\begin{figure*}
\includegraphics[width=0.49\textwidth]{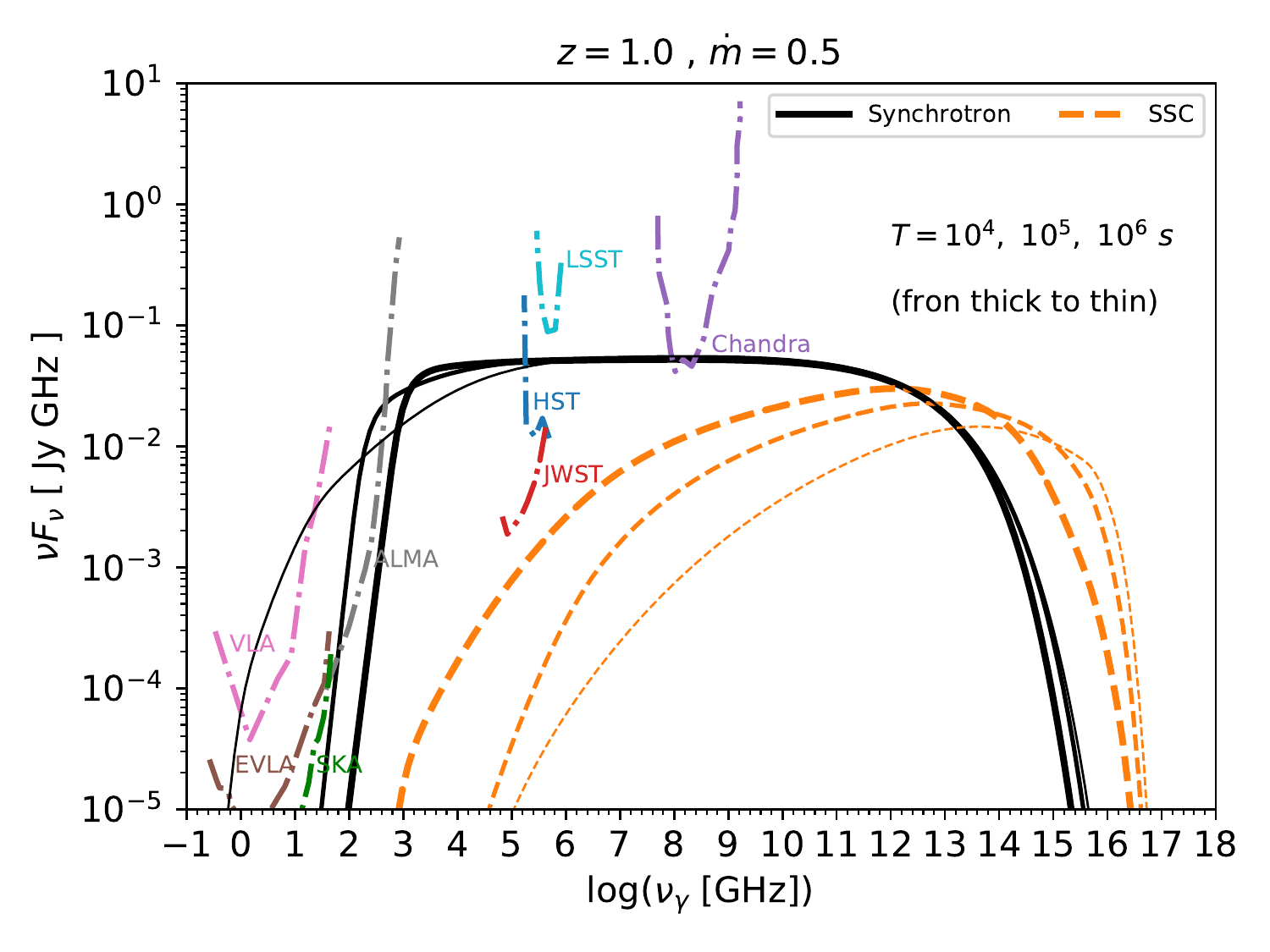}
\includegraphics[width=0.49\textwidth]{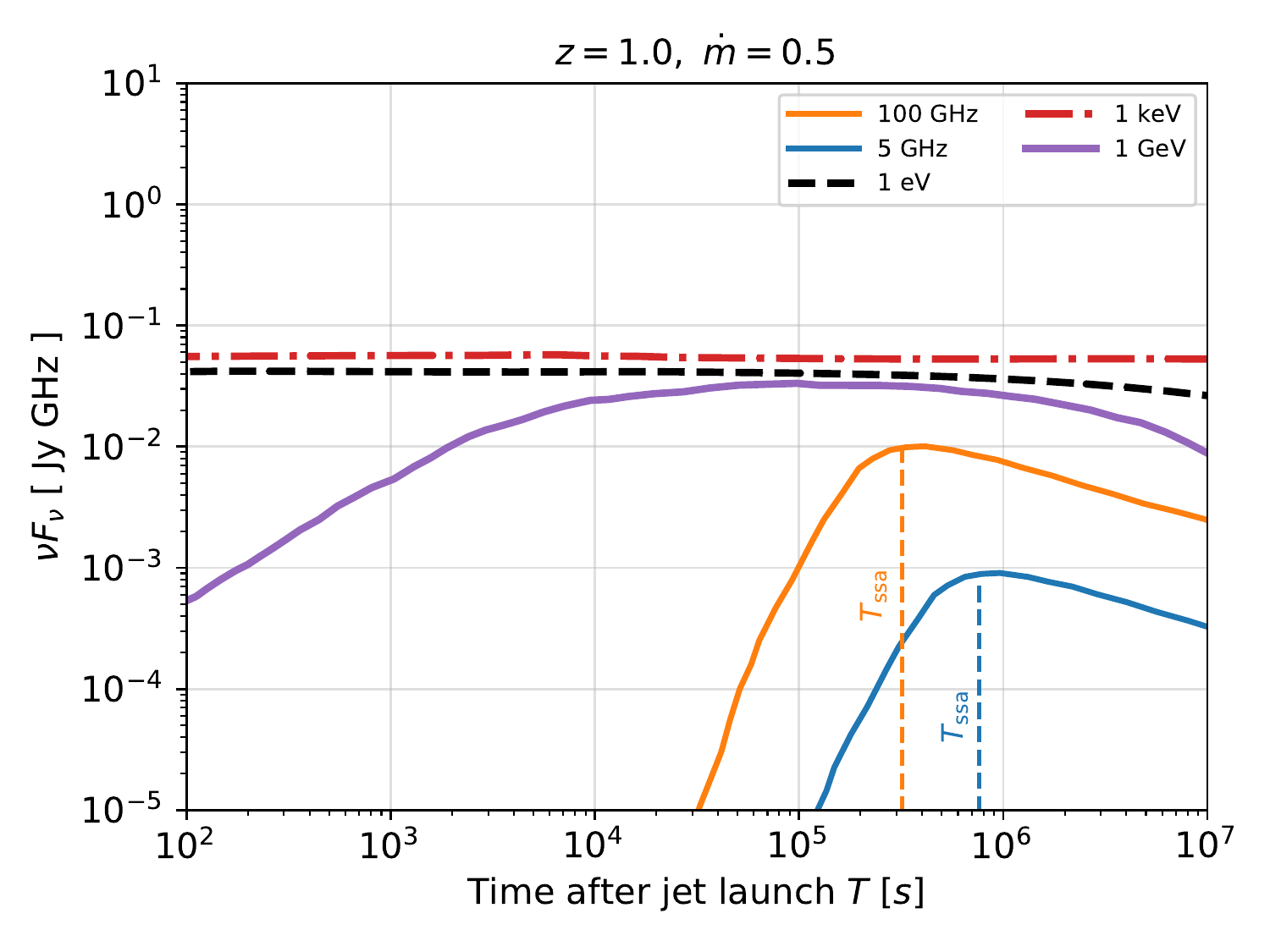}
\caption{{\bf Left panel:} Non-thermal energy spectra expected for uncollimated post-merger jets from a SMBH merger located at $z=1$. The solid and dashed lines represent the synchrotron and SSC components. 
The dash-dotted lines show the sensitivity curves for current and future detectors. {\bf Right panel:} Multi-wavelength light curves. The yellow and blue dashed vertical lines illustrate respectively the characteristic times, e.g., $T_{\rm ssa}$, of 100 GHz and 5 GHz emissions. The used parameters are $\dot m=0.5$, $M_{\rm BH}=10^6~M_\odot$, $\tilde{\eta}_w={10}^{-1.5}$, $\eta_j=1$, $\theta_j=10^{-0.5}$, $s=2.0$, $\zeta_e=0.4$, $\epsilon_e=0.1$ and $\epsilon_B=0.01$.
} 
\label{fig:spec}
\end{figure*}

\section{Electromagnetic emission from post-merger jets}\label{sec:emissions}
With the jet dynamics presented in the previous section, we calculate the EM spectra resulting from synchrotron and SSC emission. As in the standard theory of GRB afterglows~\citep[e.g.,][]{Meszaros:2006rc}, we assume that electrons are accelerated at the external forward shock with a power-law spectral index $s$. The energy fractions of the downstream energy density converted to non-thermal electron and magnetic field energy are defined as $\epsilon_e$ and $\epsilon_B$, respectively. The upstream number density is given by $n_{h,u}\approx \varrho_w(R_h)/m_p \propto R_h^{-2}$, and $B\approx[\epsilon_B 32\pi \Gamma_h(\Gamma_h-1)n_{h,u}m_pc^2]^{1/2}$ is the downstream magnetic field strength.

In the relativistic limit ($\Gamma_h\gg1$), the characteristic injection frequency $\nu_m$ and the cooling frequency $\nu_c$ in the observer frame are written respectively as,
\begin{linenomath*}
\begin{equation}\begin{split}
\nu_m&\approx\frac{3\Gamma_h\gamma_m^2eB}{4\pi(1+z) m_ec}\\
    &\simeq3.4\times10^3{\ \rm GHz}\ \epsilon_{e,-1}^2 {\zeta_{e,-0.4}^{2}}\epsilon_{B,-2}^{1/2}\eta_j^{1/2}T_4^{-1}\\
    &~~~\times(\dot m/0.5)^{1/2}M_{\rm BH,6}^{1/2}\theta_{j,-0.5}^{-1}
    \end{split}
    \label{eq:e_m}
\end{equation}
\end{linenomath*}
and
\begin{linenomath*}
\begin{equation}\begin{split}
\nu_c&\approx\frac{3\Gamma_h\gamma_c^2eB}{4\pi(1+z) m_ec}\\
    &\simeq4.6\times10^2{\ \rm GHz\ }(1+z)^{-2}(1+Y)^{-2}\epsilon_{B,-2}^{-3/2}\\
    &~~~\times\tilde\eta_{w,-1.5}^{-2}\theta_{j,-0.5}^{-1}(\dot m/0.5)^{-3/2}T_4\beta_{d,-1}^2M_{\rm BH,6}^{-3/2},
    \end{split}
    \label{eq:e_c}
\end{equation}
\end{linenomath*}
where $\gamma_m=\epsilon_e\zeta_e (\Gamma_h-1)m_p/m_e$ is the electron minimum Lorentz factor, and $\gamma_c=6\pi m_ec/[(1+Y) T'\sigma_TB^2]$ is the cooling Lorentz factor. Here, {$\zeta_e=g_s/f_e=1/[f_e\ln(\gamma_M/\gamma_m)]\sim 0.3-0.4$} is constrained by the particle-in-cell simulations \citep{park2015simultaneous} (where $f_e$ is the fraction of accelerated electrons and the maximum Lorentz factor of electrons is $\gamma_M={(6\pi e)^{1/2}/[\sigma_TB(1+Y)]^{1/2}}$), $Y$ is the Compton parameter, and $T'=\hat{T}/\Gamma_h\approx2\Gamma_h T/(1+z)$ is the comoving time. 
For example, at $T=10^4~\rm s$, we have $Y\simeq2.4$,
corresponding to the fast cooling regime. It changes to the slow cooling regime on a time scale from days to weeks. We obtain the peak synchrotron flux \citep[e.g.,][]{wijers1999physical}
\begin{linenomath*}
\begin{equation}\begin{split}
F_{\nu,\rm syn}^{\rm max}&\approx\frac{(1+z) {(0.6f_en_{h,u}R_h^3)}\Gamma_h e^3B}{\sqrt{3} m_ec^2 d_L^2}\\
&\simeq 0.24~{\rm m Jy}~(1+z)g_{s,-1.2}\zeta_{e,-0.4}^{-1}(\dot m/0.5)^{3/2}\\
&~~~\times\eta_j^{1/2}\epsilon_{B,-2}^{1/2}\tilde\eta_{w,-1.5}\beta_{d,-1}^{-1}\theta_{j,-0.5}^{-1}M_{\rm BH,6}^{3/2}d_{L,28}^{-2}.
\end{split}
\label{eq:f_max}
\end{equation}
\end{linenomath*}

The low-frequency synchrotron emission is subject to synchrotron self-absorption (SSA). The SSA optical depth is written as
$\tau_{\rm ssa}(\nu)={\xi_s}{en_{h,u}R_h}(\nu/\nu_n)^{-p}/[{B\gamma_n^5}]$,
where $\nu$ is the observed frequency, $\xi_s\sim5-10$ depends on the electron spectral index, $\gamma_n=\min\left[\gamma_m,\ \gamma_c\right]$, $\nu_n=\gamma_n^2eB/[(1+z)m_ec]$, $p=5/3$ for $\nu<\nu_n$ and $p=(4+s)/2$ or $p=3$ for $\nu>\nu_n$ depending on the slow or fast cooling regime (e.g., \citealt{panaitescu2000analytic,murase2014probing}). 
The critical time scales set by $\tau_{\rm ssa}=1$ for $\nu<\nu_n$ and for $\nu>\nu_n$ are 
$T_{\rm ssa}\simeq 5.4\times10^5~{\rm s}~\xi_{s,1}^{3/10}(1+z)^{1/2}(1+Y)^{1/2}\epsilon_{B,-2}^{3/5}\left(\frac{\nu}{1\ \rm GHz}\right)^{-1/2}\tilde\eta_{w,-1.5}^{11/10}\beta_{d,-1}^{-11/10}(\dot m/0.5)^{9/10}M_{\rm BH,6}^{9/10}$ and $T_{\rm ssa}\simeq 3.5\times10^5~{\rm s}~\xi_{s,1}^{1/2}(1+z)^{-1/2}(1+Y)^{-1/2}M_{\rm BH,6}^{1/2}\left(\frac{\nu}{100\ \rm GHz}\right)^{-3/2}(\dot m/0.5)^{1/2}M_{\rm BH,6}^{1/2}\tilde\eta_{w,-1.5}^{1/2} \beta_{d,-1}^{-1/2}$, respectively.
Thus, we expect that EM emission at 5 GHz and 100 GHz reaches a peak about a few days after the jet launch ($T_{\rm ssa}\simeq 7.1\times10^5~\rm s$ and $T_{\rm ssa}\simeq 3.1\times10^5~\rm s$, respectively, in our fiducial case with $\xi_s=8.7$).

We numerically calculate the electron distribution and the resulting synchrotron and SSC spectra of the forward shock, following the method used in \cite{murase2011implications} and \cite{Zhang:2020qbt}.
We solve the continuity equation that describes the evolution of the electron spectra and calculate the synchrotron/SSC components, in which the trans-relativistic regime can be consistently treated as in \cite{Zhang:2020qbt}. Combining the obtained radio, millimeter, optical and X-ray light curves with the sensitivities of corresponding detectors, we discuss the possibility of follow-up observations of the EM counterpart. 

The left panel of Fig.~\ref{fig:spec} shows the snapshots of synchrotron and SSC spectra at $T=10^4-10^6~\rm s$ for an on-axis source located at $z=1$. 
We assume $s=2.0$, $\epsilon_e=0.1$, and $\epsilon_B=0.01$. The solid and dashed lines correspond to the synchrotron and SSC components. Very high-energy gamma-ray emission at $\gtrsim 1$ TeV energies is suppressed due to the Klein-Nishina effect~\citep[e.g.,][]{murase2011implications,Zhang:2020qbt}, and the $\gamma\gamma$ annihilation with the extragalactic background light (EBL). For the EBL correction, $\gamma\gamma$ optical depth data from Model C in \cite{finke2010modeling} is used.
To show how the EM signal evolves with time, we illustrate the gamma-ray (1 GeV), X-ray (1 keV), UV (1 eV) and radio (5 GHz and 100 GHz) light curves in the right panel. In particular, before the characteristic time $T_{\rm ssa}$ (shown as the vertical yellow and blue lines). The radio emission is suppressed by the SSA process, which is implemented by multiplying $(1-e^{-\tau_{\rm ssa}})/\tau_{\rm ssa}$.
 
\begin{figure}
\centering
\includegraphics[width=0.5\textwidth]{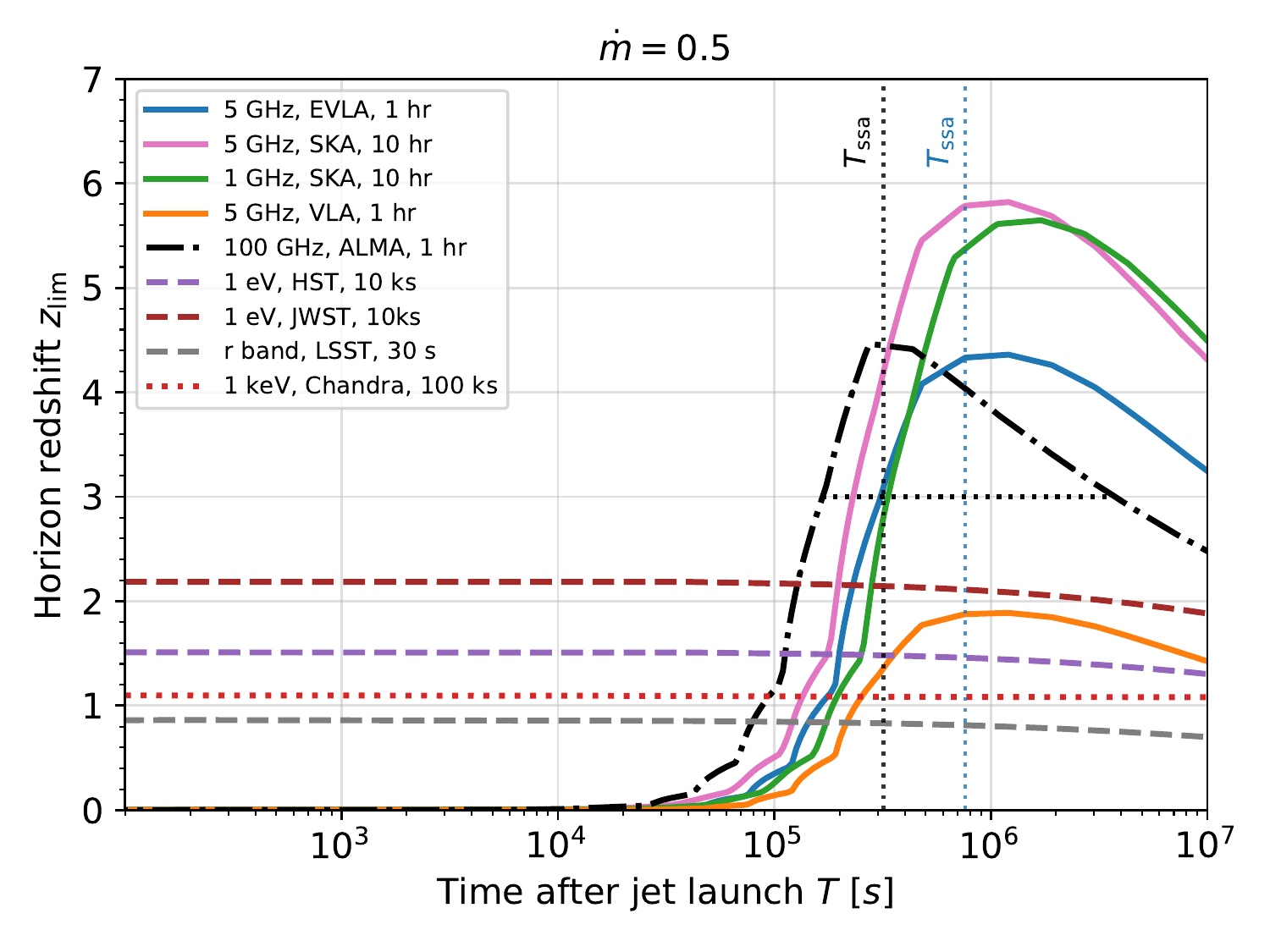}
\caption{Detection horizons for multi-wavelength detectors, e.g., SKA, VLA, EVLA, ALMA, HST, JWST, LSST and Chandra. The horizontal dotted line shows the 100 GHz detection window for ALMA assuming a source located at $z=3$. Similar to Fig. \ref{fig:spec}, the dotted vertical lines are the characteristic times of 5 GHz and 100 GHz signals.}
\label{fig:horizon}
\end{figure}

It is useful to discuss the detection horizon $d_{\rm lim}$ for some detectors such as the Square Kilometre Array (SKA), Very Large Array (VLA), Expanded VLA (EVLA), Atacama Large Millimeter Array (ALMA), Hubble Space Telescope (HST), James Webb Space Telescope (JWST), Large Synoptic Survey Telescope (LSST) and the high-resolution camera on the Chandra X-ray Observatory (Chandra) \footnote{For information on these facilities see, e.g., : VLA (\url{http://www.vla.nrao.edu}), EVLA (\url{http://www.aoc.nrao.edu/evla/}), SKA (\url{https://www.skatelescope.org}), ALMA (\url{https://public.nrao.edu/telescopes/alma/}), HST (\url{https://www.nasa.gov/mission_pages/hubble/main/index.html}), JWST (\url{https://stsci.edu/jwst}), LSST (\url{https://www.lsst.org/scientists/scibook}) and Chandra (\url{https://cxc.cfa.harvard.edu/cdo/about_chandra/})} as functions of the observation time $T$. Given the observed flux $F_\nu(\nu_\gamma,T,z)$ at the observer time $T$ from an on-axis source located at redshift $z$, the horizon can be calculated iteratively via
\begin{linenomath*}
\begin{equation}
d_{\rm lim}(\nu_\gamma,T)=d_L\left(\frac{\frac{1}{\Delta T_{\rm exp}}\int_T^{T+\Delta T_{\rm exp}}F_{\nu}(\nu_\gamma, t,z)dt}{F_{\rm lim}(\nu_\gamma,\Delta T_{\rm exp})}\right)^{1/2},
\end{equation}
\end{linenomath*}
where $F_{\rm lim}(\nu_\gamma,\Delta T_{\rm exp})$ is the detector sensitivity normalized to the exposure time $\Delta T_{\rm exp}$. For example, specifying the detection frequency $\nu=100~\rm GHz$, the sensitivity of ALMA is approximately $34~\mu\rm Jy$ for one-hour integration, e.g., $\Delta T_{\rm exp}=1$ hour. Fig. \ref{fig:horizon} indicates the detection horizons for SKA (5 GHz, $\Delta T_{\rm exp}=10$ hr), {SKA (1 GHz\footnote{{At 1 GHz, the SKA field-of-view can reach $\gtrsim$1 deg$^2$}}, $\Delta T_{\rm exp}=10$ hr)}, VLA (5 GHz, $\Delta T_{\rm exp}=1$ hr), ALMA (100 GHz, $\Delta T_{\rm exp}=1$ hr), JWST (1 eV, $\Delta T_{\rm exp}=10$ ks), HST (1 eV, $\Delta T_{\rm exp}=10$ ks), LSST (r-band, point source exposure time $\Delta T_{\rm exp}=30$ s in the 3-day revisit time), and Chandra (1 keV, $\Delta T_{\rm exp}=100$ ks). The vertical black and blue dotted lines respectively illustrate the times $T_{\rm ssa}$ at which photons at 100 GHz and 5 GHz bands start to survive from the synchrotron self-absorption.

From Fig.~\ref{fig:horizon}, we expect that ALMA, SKA and EVLA can detect SMBH mergers in the radio bands respectively out to redshifts of $z\sim4-6$. Remarkably, the optical and X-ray signals from the mergers in the range $1\lesssim z\lesssim2$ can also be identified through targeted searches by Chandra, HST and JWST in a long duration after the merger. In addition, we can estimate the observation time for each detector if the luminosity distance of the merger is specified. For example, a source located at $z=3$ would remain detectable by ALMA for roughly 20-30 days (see the black dotted horizontal line). 
One caveat is that this calculation is carried out in the ideal case where the detectors can point to the position of the source and start the observation immediately after the EM signal reaches the Earth. 
We discuss the sky coverage and a detection strategy in the following Sec. \ref{sec:discussion}.

\section{Summary and Discussion}\label{sec:discussion}
We investigated broadband non-thermal EM emission from electrons accelerated at the external forward shock expected in post-merger jets from the coalescence of SMBHs. 
In our model, the jets can be launched at $t_{\rm delay}\sim t_{\rm vis}\sim(0.003-0.1)M_{\rm BH,6}$~yr after the coalescence. The time lag is primarily determined by the scale height of the circumbinary disk and the viscosity parameter. We found that, for a moderate accretion rate ($\dot{m}\sim0.5$), the multi-wavelength emission from such a system may persist at detectable levels for months after the jet launch, depending on the facilities and the luminosity distance. Moreover, according to our model, {the sources with moderate $\dot m=0.5$ can be detected up to $z\sim5-6$, covering the range that LISA-like GW detectors have the best detection chance, e.g., $z\sim1-2$, in which $(1-10)f_b$ mergers per year are expected ~\citep{menou2001merger,enoki2004gravitational,arun2009massive,amaro2012low,2019ApJ...886..146D,2019BAAS...51c.112K}.
Here $f_b\sim1/(2\Gamma_h^2)$ is the beaming factor in our model. Because the jet head Lorentz factor is as low as $\Gamma_h\lesssim2$, the EM emission from the forward shock region is not highly beamed and we expect $f_b\sim0.1-1$.} This makes the binary SMBH mergers interesting targets for future multi-messenger studies. 
If super-Eddington accretion (e.g., $\dot m\sim10$) occurs, as was optimistically assumed in \cite{yuan2020high}, even LSST and Chandra could detect EM signals from the sources in the redshift range $4\lesssim z\lesssim6$.
We showed the case of $s=2.0$ for the demonstration. If a larger spectral index, e.g., $s\sim2.2-2.4$, is used, as expected from observations of GRB afterglows, the radio detection would be more promising whereas a higher accretion rate would be required for successful optical and X-ray observations.

The density of the premerger bubble, which was assumed to be a wind profile, is subject to large uncertainties. The extrapolation in the density distribution would be applicable up to an outer wind radius of $\sim10^{14}-10^{16}$~cm. The density predicted by equation~\ref{eq:rho_w} would drop below that of the central molecular zone {(indicated as the circumnuclear environment in Fig.\ref{fig:schematic})}, which may lead to the increase of radio emission.
In addition, a cocoon formed along with the jet, depending on uncertain details of the medium, could produce thermal photons which may not only lead to detectable signals but also serve as seed photons for inverse-Compton emission.
We focused on the more secure EM emission from the forward shock region as the jet propagates in the wind. In this sense, our prediction for the fluxes are conservative.

EM emission from the external reverse shock and internal shocks can also be expected \citep[e.g., ][for the reverse shock emission in GRBs]{meszaros1999grb,kobayashi2002grb}. 
Qualitatively, the ratio between the peak fluxes of the reverse and forward shock emission depends on the value of $\Gamma_j$, and the reverse shock contribution might be important for $\Gamma_j\gg \Gamma_h$. 

Previous studies based on general relativistic three dimensional magnetohydrodynamics simulations have shown that the circumbinary disk and the corona can emit light in UV/EUV bands \citep[e.g.,][]{d2018electromagnetic}, while X-ray and infrared emission from the post-merger circumbinary disk are expected to last for years \citep{milosavljevic2005afterglow,schnittman2008infrared}.
In the pre-merger phase, the orbits of dual SMBH cores may be identified by radio facilities such as VLBI \citep[e.g.,][]{2006ApJ...646...49R}.
Blind searches could identify radio or UV/EUV sources from the binary SMBH systems, which would provide complementary constraints on the source location, the accretion rate and the ambient gaseous environment.

Our model can provide a guidance, including the onset times and the detection windows, in developing detection strategies for future EM follow-up observations, once GW signals are detected. Considering the large uncertainties in the localization with GW detectors, an initial follow-up using large field-of-view (FOV) telescopes, like SKA and LSST, would be necessary to more precisely localize the position of the source. After that, we can use the putative positional information from the initial follow-up imaging to guide the observation of narrower FOV telescopes. {In particular, for high-redshift mergers in the range $z\sim2-5$, EM follow-up observations rely more on radio detectors, and the detection is possible a few weeks after the merger. SKA needs the source localization before follow-up observations by VLA and ALMA. On the other hand, if the merger is close enough (e.g., $z\sim1$), LISA observations staring from a few weeks before the merger can localize the merger with a median precision of $\sim$1 deg$^2$ \citep{2020PhRvD.102h4056M}. In this case, LISA and LSST can jointly guide other X-ray and optical facilities in the very early stage. Amid these two regimes, e.g., $z\sim1-2$, detections in the optical and X-ray bands using HST, JWST and Chandra would be promising if the source is localized by SKA.}
%


\begin{acknowledgements}
C.C.Y. and P.M. acknowledge support from the Eberly Foundation. The work of K.M. is supported by NSF Grant No.~AST-1908689, and KAKENHI No.~20H01901 and No.~20H05852. B.T.Z. acknowledges the IGC fellowship. S.S.K. acknowledges the JSPS Research Fellowship, JSPS KAKENHI Grant No. 19J00198.
\end{acknowledgements}
\bibliographystyle{aasjournal}
\bibliography{ref}

\begin{thebibliography}{}
\expandafter\ifx\csname natexlab\endcsname\relax\def\natexlab#1{#1}\fi
\providecommand{\url}[1]{\href{#1}{#1}}
\providecommand{\dodoi}[1]{doi:~\href{http://doi.org/#1}{\nolinkurl{#1}}}
\providecommand{\doeprint}[1]{\href{http://ascl.net/#1}{\nolinkurl{http://ascl.net/#1}}}
\providecommand{\doarXiv}[1]{\href{https://arxiv.org/abs/#1}{\nolinkurl{https://arxiv.org/abs/#1}}}

\bibitem[{Akiyama {et~al.}(2019)Akiyama, Alberdi, Alef, Asada, Azulay, Baczko,
  Ball, Balokovi{\'c}, Barrett, Bintley, {et~al.}}]{akiyama2019first}
Akiyama, K., Alberdi, A., Alef, W., {et~al.} 2019, \apjl, 875, L5,
  \dodoi{10.3847/2041-8213/ab0f43}

\bibitem[{Amaro-Seoane {et~al.}(2012)Amaro-Seoane, Aoudia, Babak, Binetruy,
  Berti, Bohe, Caprini, Colpi, Cornish, Danzmann, {et~al.}}]{amaro2012low}
Amaro-Seoane, P., Aoudia, S., Babak, S., {et~al.} 2012, Classical and Quantum
  Gravity, 29, 124016, \dodoi{10.1088/0264-9381/29/12/124016}

\bibitem[{Amaro-Seoane {et~al.}(2017)Amaro-Seoane, Audley, Babak, Baker,
  Barausse, Bender, Berti, Binetruy, Born, Bortoluzzi,
  {et~al.}}]{amaro2017laser}
Amaro-Seoane, P., Audley, H., Babak, S., {et~al.} 2017, arXiv:1702.00786.
\newblock \doarXiv{1702.00786}

\bibitem[{Arun {et~al.}(2009)Arun, Babak, Berti, Cornish, Cutler, Gair, Hughes,
  Iyer, Lang, Mandel, {et~al.}}]{arun2009massive}
Arun, K., Babak, S., Berti, E., {et~al.} 2009, Classical and Quantum Gravity,
  26, 094027, \dodoi{10.1088/0264-9381/26/9/094027}

\bibitem[{{Arzoumanian} {et~al.}(2020){Arzoumanian}, {Baker}, {Blumer},
  {B{\'e}csy}, {Brazier}, {Brook}, {Burke-Spolaor}, {Chatterjee}, {Chen},
  {Cordes}, {Cornish}, {Crawford}, {Cromartie}, {Decesar}, {Demorest}, {Dolch},
  {Ellis}, {Ferrara}, {Fiore}, {Fonseca}, {Garver-Daniels}, {Gentile}, {Good},
  {Hazboun}, {Holgado}, {Islo}, {Jennings}, {Jones}, {Kaiser}, {Kaplan},
  {Kelley}, {Key}, {Laal}, {Lam}, {Lazio}, {Lorimer}, {Luo}, {Lynch},
  {Madison}, {McLaughlin}, {Mingarelli}, {Ng}, {Nice}, {Pennucci}, {Pol},
  {Ransom}, {Ray}, {Shapiro-Albert}, {Siemens}, {Simon}, {Spiewak}, {Stairs},
  {Stinebring}, {Stovall}, {Sun}, {Swiggum}, {Taylor}, {Turner}, {Vallisneri},
  {Vigeland}, {Witt}, \& {Nanograv Collaboration}}]{2020ApJ...905L..34A}
{Arzoumanian}, Z., {Baker}, P.~T., {Blumer}, H., {et~al.} 2020, \apjl, 905,
  L34, \dodoi{10.3847/2041-8213/abd401}

\bibitem[{Baker {et~al.}(2019)Baker, Barke, Bender, Berti, Caldwell, Conklin,
  Cornish, Ferrara, Holley-Bockelmann, Kamai, {et~al.}}]{baker2019space}
Baker, J., Barke, S.~F., Bender, P.~L., {et~al.} 2019, 51, 243.
\newblock \doarXiv{1907.11305}

\bibitem[{Begelman {et~al.}(1980)Begelman, Blandford, \&
  Rees}]{begelman1980massive}
Begelman, M.~C., Blandford, R.~D., \& Rees, M.~J. 1980, \nat, 287, 307,
  \dodoi{10.1038/287307a0}

\bibitem[{Blandford \& Znajek(1977)}]{blandford1977electromagnetic}
Blandford, R.~D., \& Znajek, R.~L. 1977, \mnras, 179, 433,
  \dodoi{10.1093/mnras/179.3.433}

\bibitem[{Bromberg {et~al.}(2011)Bromberg, Nakar, Piran,
  {et~al.}}]{bromberg2011propagation}
Bromberg, O., Nakar, E., Piran, T., {et~al.} 2011, \apj, 740, 100,
  \dodoi{10.1088/0004-637X/740/2/100}

\bibitem[{{Dal Canton} {et~al.}(2019){Dal Canton}, {Mangiagli}, {Noble},
  {Schnittman}, {Ptak}, {Klein}, {Sesana}, \& {Camp}}]{2019ApJ...886..146D}
{Dal Canton}, T., {Mangiagli}, A., {Noble}, S.~C., {et~al.} 2019, \apj, 886,
  146, \dodoi{10.3847/1538-4357/ab505a}

\bibitem[{d’Ascoli {et~al.}(2018)d’Ascoli, Noble, Bowen, Campanelli,
  Krolik, \& Mewes}]{d2018electromagnetic}
d’Ascoli, S., Noble, S.~C., Bowen, D.~B., {et~al.} 2018, \apj, 865, 140,
  \dodoi{10.3847/1538-4357/aad8b4}

\bibitem[{Enoki {et~al.}(2004)Enoki, Inoue, Nagashima, \&
  Sugiyama}]{enoki2004gravitational}
Enoki, M., Inoue, K.~T., Nagashima, M., \& Sugiyama, N. 2004, \apj, 615, 19,
  \dodoi{10.1086/424475}

\bibitem[{Farris {et~al.}(2015)Farris, Duffell, MacFadyen, \&
  Haiman}]{farris2015binary}
Farris, B.~D., Duffell, P., MacFadyen, A.~I., \& Haiman, Z. 2015, \mnras, 447,
  L80, \dodoi{10.1093/mnrasl/slu184}

\bibitem[{Finke {et~al.}(2010)Finke, Razzaque, \& Dermer}]{finke2010modeling}
Finke, J.~D., Razzaque, S., \& Dermer, C.~D. 2010, \apj, 712, 238,
  \dodoi{10.1088/0004-637X/712/1/238}

\bibitem[{Haiman(2017)}]{haiman2017electromagnetic}
Haiman, Z. 2017, \prd, 96, 023004, \dodoi{10.1103/PhysRevD.96.023004}

\bibitem[{Haiman {et~al.}(2009)Haiman, Kocsis, Menou, Lippai, \&
  Frei}]{haiman2009identifying}
Haiman, Z., Kocsis, B., Menou, K., Lippai, Z., \& Frei, Z. 2009, Classical and
  Quantum Gravity, 26, 094032, \dodoi{10.1088/0264-9381/26/9/094032}

\bibitem[{Jiang {et~al.}(2019{\natexlab{a}})Jiang, Blaes, Stone, \&
  Davis}]{jiang2019global}
Jiang, Y.-F., Blaes, O., Stone, J.~M., \& Davis, S.~W. 2019{\natexlab{a}},
  \apj, 885, 144, \dodoi{10.3847/1538-4357/ab4a00}

\bibitem[{Jiang {et~al.}(2019{\natexlab{b}})Jiang, Stone, \&
  Davis}]{jiang2019super}
Jiang, Y.-F., Stone, J.~M., \& Davis, S.~W. 2019{\natexlab{b}}, \apj, 880, 67,
  \dodoi{10.3847/1538-4357/ab29ff}

\bibitem[{{Kara} {et~al.}(2019){Kara}, {Margutti}, {Keivani}, {Fong}, {Cenko},
  {Noble}, {Mushotzky}, {Burns}, {Ryan}, {Ruan}, {Haggard}, {Burrows}, {Fox},
  \& {Caputo}}]{2019BAAS...51c.112K}
{Kara}, E., {Margutti}, R., {Keivani}, A., {et~al.} 2019, \baas, 51, 112.
\newblock \doarXiv{1903.05287}

\bibitem[{{Kelly} {et~al.}(2017){Kelly}, {Baker}, {Etienne}, {Giacomazzo}, \&
  {Schnittman}}]{2017PhRvD..96l3003K}
{Kelly}, B.~J., {Baker}, J.~G., {Etienne}, Z.~B., {Giacomazzo}, B., \&
  {Schnittman}, J. 2017, \prd, 96, 123003, \dodoi{10.1103/PhysRevD.96.123003}

\bibitem[{Kobayashi \& Zhang(2003)}]{kobayashi2002grb}
Kobayashi, S., \& Zhang, B. 2003, \apjl, 582, L75, \dodoi{10.1086/367691}

\bibitem[{Kormendy \& Ho(2013)}]{kormendy2013coevolution}
Kormendy, J., \& Ho, L.~C. 2013, \araa, 51, 511,
  \dodoi{10.1146/annurev-astro-082708-101811}

\bibitem[{{Kroupa} {et~al.}(2020){Kroupa}, {Subr}, {Jerabkova}, \&
  {Wang}}]{2020MNRAS.498.5652K}
{Kroupa}, P., {Subr}, L., {Jerabkova}, T., \& {Wang}, L. 2020, \mnras, 498,
  5652, \dodoi{10.1093/mnras/staa2276}

\bibitem[{{Mangiagli} {et~al.}(2020){Mangiagli}, {Klein}, {Bonetti}, {Katz},
  {Sesana}, {Volonteri}, {Colpi}, {Marsat}, \& {Babak}}]{2020PhRvD.102h4056M}
{Mangiagli}, A., {Klein}, A., {Bonetti}, M., {et~al.} 2020, \prd, 102, 084056,
  \dodoi{10.1103/PhysRevD.102.084056}

\bibitem[{Menou {et~al.}(2001)Menou, Haiman, \& Narayanan}]{menou2001merger}
Menou, K., Haiman, Z., \& Narayanan, V.~K. 2001, \apj, 558, 535,
  \dodoi{10.1086/322310}

\bibitem[{M\'esz\'aros(2006)}]{Meszaros:2006rc}
M\'esz\'aros, P. 2006, Rept. Prog. Phys., 69, 2259,
  \dodoi{10.1088/0034-4885/69/8/R01}

\bibitem[{M{\'e}sz{\'a}ros {et~al.}(2019)M{\'e}sz{\'a}ros, Fox, Hanna, \&
  Murase}]{meszaros2019multi}
M{\'e}sz{\'a}ros, P., Fox, D.~B., Hanna, C., \& Murase, K. 2019, Nature Reviews
  Physics, 1, 585, \dodoi{10.1038/s42254-019-0101-z}

\bibitem[{M{\'e}sz{\'a}ros \& Rees(1999)}]{meszaros1999grb}
M{\'e}sz{\'a}ros, P., \& Rees, M.~J. 1999, \mnras, 306, L39,
  \dodoi{10.1046/j.1365-8711.1999.02800.x}

\bibitem[{Milosavljevi{\'c} \& Phinney(2005)}]{milosavljevic2005afterglow}
Milosavljevi{\'c}, M., \& Phinney, E.~S. 2005, \apjl, 622, L93,
  \dodoi{10.1086/429618}

\bibitem[{{Mingarelli} {et~al.}(2017){Mingarelli}, {Lazio}, {Sesana}, {Greene},
  {Ellis}, {Ma}, {Croft}, {Burke-Spolaor}, \& {Taylor}}]{2017NatAs...1..886M}
{Mingarelli}, C. M.~F., {Lazio}, T. J.~W., {Sesana}, A., {et~al.} 2017, Nature
  Astronomy, 1, 886, \dodoi{10.1038/s41550-017-0299-6}

\bibitem[{Mizuta \& Ioka(2013)}]{mizuta2013opening}
Mizuta, A., \& Ioka, K. 2013, \apj, 777, 162,
  \dodoi{10.1088/0004-637X/777/2/162}

\bibitem[{{Moesta} {et~al.}(2012){Moesta}, {Alic}, {Rezzolla}, {Zanotti}, \&
  {Palenzuela}}]{2012ApJ...749L..32M}
{Moesta}, P., {Alic}, D., {Rezzolla}, L., {Zanotti}, O., \& {Palenzuela}, C.
  2012, \apjl, 749, L32, \dodoi{10.1088/2041-8205/749/2/L32}

\bibitem[{Murase \& Bartos(2019)}]{Murase:2019tjj}
Murase, K., \& Bartos, I. 2019, Ann. Rev. Nucl. Part. Sci., 69, 477,
  \dodoi{10.1146/annurev-nucl-101918-023510}

\bibitem[{Murase \& Ioka(2013)}]{Murase:2013ffa}
Murase, K., \& Ioka, K. 2013, Phys. Rev. Lett., 111, 121102,
  \dodoi{10.1103/PhysRevLett.111.121102}

\bibitem[{Murase {et~al.}(2014)Murase, Thompson, \& Ofek}]{murase2014probing}
Murase, K., Thompson, T.~A., \& Ofek, E.~O. 2014, \mnras, 440, 2528,
  \dodoi{10.1093/mnras/stu384}

\bibitem[{Murase {et~al.}(2011)Murase, Toma, Yamazaki, \&
  M{\'e}sz{\'a}ros}]{murase2011implications}
Murase, K., Toma, K., Yamazaki, R., \& M{\'e}sz{\'a}ros, P. 2011, \apj, 732,
  77, \dodoi{10.1088/0004-637X/732/2/77}

\bibitem[{Nakar(2015)}]{Nakar:2015tma}
Nakar, E. 2015, Astrophys. J., 807, 172, \dodoi{10.1088/0004-637X/807/2/172}

\bibitem[{Ohsuga {et~al.}(2009)Ohsuga, Mineshige, Mori, \&
  Kato}]{ohsuga2009global}
Ohsuga, K., Mineshige, S., Mori, M., \& Kato, Y. 2009, \pasj, 61, L7,
  \dodoi{10.1093/pasj/61.3.L7}

\bibitem[{{Panaitescu} \& {Kumar}(2000)}]{panaitescu2000analytic}
{Panaitescu}, A., \& {Kumar}, P. 2000, \apj, 543, 66, \dodoi{10.1086/317090}

\bibitem[{Park {et~al.}(2015)Park, Caprioli, \&
  Spitkovsky}]{park2015simultaneous}
Park, J., Caprioli, D., \& Spitkovsky, A. 2015, \prl, 114, 085003,
  \dodoi{10.1103/PhysRevLett.114.085003}

\bibitem[{Pringle(1981)}]{pringle1981accretion}
Pringle, J. 1981, \araa, 19, 137, \dodoi{10.1146/annurev.aa.19.090181.001033}

\bibitem[{{Ravi}(2018)}]{2018ASPC..517..781R}
{Ravi}, V. 2018, in Astronomical Society of the Pacific Conference Series, Vol.
  517, Science with a Next Generation Very Large Array, ed. E.~{Murphy}, 781

\bibitem[{{Rodriguez} {et~al.}(2006){Rodriguez}, {Taylor}, {Zavala}, {Peck},
  {Pollack}, \& {Romani}}]{2006ApJ...646...49R}
{Rodriguez}, C., {Taylor}, G.~B., {Zavala}, R.~T., {et~al.} 2006, \apj, 646,
  49, \dodoi{10.1086/504825}

\bibitem[{{Schnittman}(2011)}]{2011CQGra..28i4021S}
{Schnittman}, J.~D. 2011, Classical and Quantum Gravity, 28, 094021,
  \dodoi{10.1088/0264-9381/28/9/094021}

\bibitem[{Schnittman \& Krolik(2008)}]{schnittman2008infrared}
Schnittman, J.~D., \& Krolik, J.~H. 2008, \apj, 684, 835,
  \dodoi{10.1086/590363}

\bibitem[{Senno {et~al.}(2016)Senno, Murase, \& M\'esz\'aros}]{Senno:2015tsn}
Senno, N., Murase, K., \& M\'esz\'aros, P. 2016, Phys. Rev. D, 93, 083003,
  \dodoi{10.1103/PhysRevD.93.083003}

\bibitem[{{Sesana} {et~al.}(2004){Sesana}, {Haardt}, {Madau}, \&
  {Volonteri}}]{2004ApJ...611..623S}
{Sesana}, A., {Haardt}, F., {Madau}, P., \& {Volonteri}, M. 2004, \apj, 611,
  623, \dodoi{10.1086/422185}

\bibitem[{Shapiro \& Teukolsky(1983)}]{shapiro2008black}
Shapiro, S.~L., \& Teukolsky, S.~A. 1983, Black holes, white dwarfs, and
  neutron stars: The physics of compact objects

\bibitem[{Tanaka \& Menou(2010)}]{tanaka2010time}
Tanaka, T., \& Menou, K. 2010, \apj, 714, 404,
  \dodoi{10.1088/0004-637X/714/1/404}

\bibitem[{Taylor {et~al.}(2019)Taylor, Burke-Spolaor, Baker, Charisi, Islo,
  Kelley, Madison, Simon, \& Vigeland}]{taylor2019supermassive}
Taylor, S.~R., Burke-Spolaor, S., Baker, P.~T., {et~al.} 2019, \baas, 51, 336.
\newblock \doarXiv{1903.08183}

\bibitem[{Tchekhovskoy {et~al.}(2011)Tchekhovskoy, Narayan, \&
  McKinney}]{tchekhovskoy2011efficient}
Tchekhovskoy, A., Narayan, R., \& McKinney, J.~C. 2011, \mnras, 418, L79,
  \dodoi{10.1111/j.1745-3933.2011.01147.x}

\bibitem[{Thorne \& Braginskii(1976)}]{thorne1976gravitational}
Thorne, K.~S., \& Braginskii, V. 1976, \apjl, 204, L1, \dodoi{10.1086/182042}

\bibitem[{Wijers \& Galama(1999)}]{wijers1999physical}
Wijers, R., \& Galama, T. 1999, \apj, 523, 177, \dodoi{10.1086/307705}

\bibitem[{Yuan {et~al.}(2020)Yuan, Murase, Kimura, \&
  M\'esz\'aros}]{yuan2020high}
Yuan, C., Murase, K., Kimura, S.~S., \& M\'esz\'aros, P. 2020, Phys. Rev. D,
  102, 083013, \dodoi{10.1103/PhysRevD.102.083013}

\bibitem[{Zhang {et~al.}(2020)Zhang, Murase, Veres, \&
  M{\'{e}}sz{\'{a}}ros}]{Zhang:2020qbt}
Zhang, B.~T., Murase, K., Veres, P., \& M{\'{e}}sz{\'{a}}ros, P. 2020, arXiv:
  2012.07796.
\newblock \doarXiv{2012.07796}

\end{thebibliography}
\end{document}